\documentclass{aa}  

\usepackage{graphicx}
\usepackage{txfonts}
\usepackage{subcaption}
\usepackage{caption}
\usepackage[colorlinks=true, citecolor=blue]{hyperref}
\usepackage[table]{xcolor}
\usepackage{ulem}

\begin{document} 


\title{Testing the sources of the peculiar abundances in globular clusters}

   \author{R. J. Vaca
          \inst{1,2}
          \and
          I. Cabrera-Ziri\inst{3}
          \and
          G. Magris C.\inst{2}
          \and
          N. Bastian\inst{4,5}
          \and
          M. Salaris\inst{6,7}
          }

   \institute{Universidad Simón Bolívar,
            Sartenejas, Baruta, Edo. Miranda - Apartado 89000 Cable Unibolivar, Caracas, Venezuela\\
            \email{14-11106@usb.ve}
        \and
            Centro de Investigaciones de Astronomía, Av. Alberto Carnevali, Edif, CIDA, Mérida 5101, Venezuela
        \and
            Astronomisches Rechen-Institut, Zentrum für Astronomie der Universität Heidelberg, Mönchhofstraße 12-14, D-69120 Heidelberg, Germany
        \and
            Donostia International Physics Center (DIPC), Paseo Manuel de Lardizabal, 4, 20018 Donostia-San Sebastián, Guipuzkoa, Spain
        \and
            IKERBASQUE, Basque Foundation for Science, 48013 Bilbao, Spain
        \and
            Astrophysics Research Institute, Liverpool John Moores University, 146 Brownlow Hill, Liverpool L3 5RF, UK
        \and
        	INAF - Osservatorio Astronomico d'Abruzzo, Via Mentore Maggini, 64100 Teramo, Italy
            }
   \date{}

 
  \abstract
  {This work aims to analyze some of the polluters proposed in the self-enrichment scenarios put forward to explain the multiple populations in globular clusters (GCs), extending previous studies. Three scenarios with different polluter stars were tested: asymptotic giant branch stars (AGBs), high-mass interacting binaries (IBs), and fast rotating massive stars (FRMSs). With abundance data available from the Apache Point Observatory Galactic Evolution Experiment (APOGEE) survey and $\Delta Y$ estimates from precise Hubble Space Telescope (HST) photometry, twenty-six clusters were studied, increasing the number of clusters in previous studies by more than a factor of three. We also included the study of the abundances of N, C, Mg, and Al, extending previous studies that mainly focused on the abundances of He, O, and Na. In addition, we constructed an empirical model to test whether one could explain the chemical signatures of the "enriched" population of GC stars with a fixed source and dilution process based on empirical data. In agreement with work by other authors, we found that the proposed polluters can generally predict the qualitative abundance patterns in GC stars and in some cases quantitatively predict some elements, but in most cases when we compare the model yields with the observations, we find that they cannot explain the entire set of observed abundance patterns. The empirical model succeeds in reproducing the abundances of Al for a given $\Delta Y$ (and vice versa), showing that there is a direct relationship between Al and He, with one increasing proportionally to the other. However, the empirical model fails to reproduce the observed abundances of Na and N, in agreement with the results of previous works. The observed decoupling between the maximum abundances of CNO-cycle elements such as N and Na with those of Al and He provides new information and constraints for future models and could take us a step closer to understanding the origin of the peculiar abundance variations of GC stars.}


\maketitle



\section{Introduction}
Historically, globular clusters (GCs) have been thought of as simple stellar populations and, due to their old age and low metallicity, are excellent tracers of the evolution of their parent galaxy \citep[e.g.,][]{Vandenberg2013}.
However, studies have shown that the stars belonging to individual clusters can be divided into different populations with different chemical abundances while maintaining a very small age difference \citep[e.g.,][]{Cohen1978, Martocchia2018}.
Population 1 (P1) stars have the same abundances as field stars of the same metallicity and Population 2 (P2) stars have peculiar abundances\footnote{In the literature these are also referred to as 1st and 2nd generation. However, this notation implicitly suggests a specific formation mechanism that is not common to all proposed scenarios. Therefore, in this paper we use a more inclusive terminology.}.
The study of this phenomenon has advanced in the last decades thanks to the Very Large Telescope (VLT) and Hubble Space Telescope (HST) surveys.
Through these observations, studies have found that individual stars of almost all GCs show variations in the abundances of certain elements, such as He, Na, O, N, C, Al, and sometimes Mg {\citep[for a more complete discussion see][]{Bastian2018}}.
These variations are present even in main sequence stars, suggesting that they are present during the formation of the star rather than being a product of stellar evolution \citep[e.g.,][]{Gratton2019}.

Despite decades of effort, no satisfactory explanation has yet been found to describe the mechanism responsible for the observed chemical patterns \citep[e.g.,][]{Bastian2018,Gratton2019,Cassisi2020}.
Finding the origin of these variations remains a major challenge in stellar and galactic astrophysics, and the origin of this phenomenon is still an open question.

Many studies have mainly focused on Na and O \citep[e.g.,][]{Carretta2009a,Carretta2009b}. The abundances of these elements are simultaneously increased or decreased, respectively, compared to field stars of the same metallicity.
GCs also present anticorrelations between N and C, and occasionally Al and Mg \citep[e.g.,][]{Bastian2018}.
Since the observed chemical patterns are similar to the abundances produced through the CNO cycle \citep[e.g.,][]{Kudryashov1988}, most of the proposed scenarios rely on this process to explain the phenomenon.
However, GCs in our Galaxy are made up of low-mass stars that cannot process material through the CNO cycle.
For this reason, it has been suggested that the chemical patterns observed in GCs are the product of material processed in massive stars of an earlier generation \citep[e.g.,][]{Gratton2012}.

Most GCs show variations in their chemical abundances regardless of their mass or metallicity \citep[e.g.,][]{Carretta2010}.
The most developed models proposed to explain this phenomenon use a ``self enrichment'' process.
This consists of massive stars, called polluters, enriching the intracluster medium from which subsequent generations of stars are formed \citep[e.g.,][]{Gratton2012,Bastian2018}.
These polluters must meet the following criteria: they must be massive enough stars capable of burning hydrogen through the CNO cycle in their high-temperature cores (or in the case of asymptotic giant branch stars, a shell surrounding the core) and they must have short lifetimes  so as not to affect the range of ages observed in the cluster \citep[e.g.,][]{Charbonnel2016}.
Finally, the polluting stars must have chemical abundances similar to those of field stars of the same metallicity.
Their stellar ejecta pollute the medium, eventually producing stars with peculiar abundances in some elements.
The two populations are thus obtained \citep[e.g.,][]{Gratton2019}.

A recent study by \citet{Lahen2024} found through dwarf galaxy simulations that GCs forming in low metallicity star bursts of a merger could produce the multiple populations found in GCs.
They found that massive ($\gtrsim 9 M_\sun$) and very massive ($\gtrsim 100 M_\sun$) stars can acrete material into the intracluster with similar relations as the abundance patterns seen in GCs (i.e., they used massive and very massive stars as polluters).
However, once they considered the dilution with the interstellar medium needed to form new stars, they could not reproduce the observed spread in chemical composition seen in GCs.
A proposed solution to this is to consider different types of polluters, such as interacting binaries or fast rotating massive stars, that can release ten times more material into the intracluster medium.

One way to test the basic principals of self-enrichment is to compare the theoretical yields of chemical elements proposed by different polluter models with the observed abundance patterns in GCs.
However, there are few studies that have conducted a comprehensive analysis of pollutants that include multiple clusters and elements.
Many studies focus on studying pollutants with only one cluster \citep[e.g.,][]{DAntona2016,Prantzos2017}.
Following the "dilution models" introduced by \citet{Prantzos2006}, in which the material emitted by polluters is diluted by uncontaminated or primordial material, in this paper we compare the observed range of abundance measurements with the predictions of different polluters.

Previously, \citet{Bastian2015} carried out a study of different potential polluters. 
They focused on polluters with scenarios that have the most developed theoretical models: massive asymptotic giant branch stars (AGBs: $>5-9 M_\sun$) \citep[e.g.,][]{Cottrell1981,Karakas2014}, interacting massive binaries (IBs: $>10-20 M_\sun$) \citep[e.g.,][]{Grundstrom2007,deMink2009}, and fast rotating massive stars (FRMS: $\geq 20 M_\sun$) \citep[e.g.,][]{Maeder2006,Decressin2007a,Decressin2007b}.
The theoretical yields for each scenario were taken from \citet{Ventura2013}, \citet{deMink2009}, and \citet{Decressin2007a,Decressin2007b} respectively.
In these scenarios, these stars are born in the cluster and, through internal processes unique to each type of star, release enriched material and contaminate the environment.
The next stars to be born in the cluster will have the peculiar abundances.
As more stars are formed, the amount of polluting material is progressively depleted and diluted.

\begin{table}\centering
	\caption{Observations used in the study.}
	\label{tab:obs}
	\begin{tabular}{lccccc}
		\hline
		Cluster & $\Delta Y$ & [Fe/H] & $\Delta$[Al/Fe] & $\Delta$[N/Fe] & $\Delta$[Na/Fe] \\
		\hline
		104  & $0.049$ & $-0.8$ & $0.23 $ & $1.11$ & $0.53$ \\
		1851 & $0.025$ & $-1.1$ & $0.39 $ & $1.21$ & $ -  $ \\
		2808 & $0.124$ & $-1.1$ & $0.99 $ & $1.07$ & $0.44$ \\
		288  & $0.016$ & $-1.3$ & $0.19 $ & $0.87$ & $0.65$ \\
		3201 & $0.028$ & $-1.4$ & $0.48 $ & $0.97$ & $0.51$ \\
		362  & $0.026$ & $-1.1$ & $0.28 $ & $1.03$ & $ -  $ \\
		4590 & $0.012$ & $-2.2$ & $0.61 $ & $0.86$ & $0.47$ \\
		5024 & $0.044$ & $-1.9$ & $0.65 $ & $0.94$ & $ -  $ \\
		5053 & $0.004$ & $-2.2$ & $0.68 $ & $0.82$ & $ -  $ \\
		5272 & $0.041$ & $-1.4$ & $0.54 $ & $0.89$ & $ -  $ \\
		5466 & $0.007$ & $-1.8$ & $0.16 $ & $0.89$ & $ -  $ \\
		5904 & $0.037$ & $-1.2$ & $0.62 $ & $1.11$ & $0.52$ \\
		6121 & $0.014$ & $-1.1$ & $0.32 $ & $1.05$ & $0.46$ \\
		6171 & $0.024$ & $-1.0$ & $0.24 $ & $1.27$ & $0.52$ \\
		6205 & $0.052$ & $-1.5$ & $1.01 $ & $1.11$ & $ -  $ \\
		6218 & $0.011$ & $-1.3$ & $0.23 $ & $1.03$ & $0.57$ \\
		6254 & $0.029$ & $-1.5$ & $0.92 $ & $1.04$ & $0.54$ \\
		6341 & $0.039$ & $-2.2$ & $0.68 $ & $1.83$ & $ -  $ \\
		6388 & $0.067$ & $-0.5$ & $0.6  $ & $1.14$ & $0.65$ \\
		6397 & $0.008$ & $-2.0$ & $0.66 $ & $1.38$ & $0.39$ \\
		6441 & $0.081$ & $-0.4$ & $0.53 $ & $1.02$ & $ -  $ \\
		6656 & $0.041$ & $-1.7$ & $0.81 $ & $1.28$ & $ -  $ \\
		6809 & $0.026$ & $-1.8$ & $0.81 $ & $1.32$ & $0.71$ \\
		6838 & $0.024$ & $-0.7$ & $-0.03$ & $0.9 $ & $0.39$ \\
		7078 & $0.069$ & $-2.2$ & $0.74 $ & $1.66$ & $0.62$ \\
		7089 & $0.052$ & $-1.4$ & $0.71 $ & $0.82$ & $ -  $ \\
		\hline
	\end{tabular}
	\tablefoot{
		$\Delta$[N/Fe], $\Delta$[Al/Fe], and the average [Fe/H] values are taken from the APOGEE survey \citep{Abdurrouf2022}, $\Delta$[Na/Fe] are taken from \citet{Carretta2009a,Carretta2009b}, and $\Delta Y$ (the maximum spread of He) is taken from \citet{Milone2018}. $\Delta$[N/Fe], $\Delta$[Al/Fe], and $\Delta$[Na/Fe] were calculated using equation \eqref{eq:delta}, and they are further discussed in Section 5.
	}
	\label{tab:data}
\end{table}

It is important to note that all three of these models require  the ejected material to be diluted by primordial material with the chemical composition of P1, since the ejecta from the P2 stars alone cannot account for the full range of abundances observed in a given cluster. For example, there is a correlation between Na and O in the winds of the proposed AGB stars, the opposite of what has been observed \citep[e.g.,][]{Conroy2010,DErcole2011,Renzini2015}.
This will be discussed in more detail in Section 4.1.

\citet{Bastian2015} concluded that while these scenarios (AGB, IB, and FRMS) predict the qualitative behavior of some abundance patterns of GC stars, the proposed sources could not explain the  abundance variations quantitatively.
More specifically, they showed that a high degree of stochasticity is needed to explain the phenomenon across all clusters. They also found that the polluters produced too much He for a given spread in Na and O, and that these spreads are not directly correlated with the spread in He.
However, based on the data available at that time, only eight clusters and three elements were studied.
This study aims to use more recent, publicly available data to extend the original study and analyze the proposed sources of these peculiar abundances for 26 clusters and seven elements, He, C, N, O, Al, Mg, and Na.
To this end, we test the basic yields proposed in the different scenarios, excluding their other aspects (e.g. the origin of the primordial material, the mass budget problem, etc.).

This paper is divided into the following sections:
Section 2, where we present the observational data used, Section 3, where we explain the models used and discuss the different enrichment scenarios. In Sections 4 and 5, we present and discuss our results, and Section 6, where we present our conclusions.

\section{Data}

For our sample of GCs, the measurements for [N/Fe], [C/Fe], [Al/Fe], and [Mg/Fe], together with the metallicity (we use [Fe/H] as a proxy), were taken from the Apache Point Observatory Galactic Evolution Experiment (APOGEE) survey \citep{Abdurrouf2022}, while the [Na/Fe] and [O/Fe] measurements were taken from work of \citet{Carretta2009a,Carretta2009b}. These catalogs were cross-matched with the sample of red giant branch (RGB) GC stars used in \citet{Leitinger2023}, which were identified as cluster members on the basis of their proper motion, parallaxes, color and magnitudes.
The dynamical masses for the clusters were taken from \citet{Baumgardt2023}.

For He, we used the measurements available in \citet{Milone2018} of the maximum spread of He ($\Delta Y$) in a given cluster.
This value adds an additional constraint to the models that must be taken into consideration (i.e., the model must be able to predict the observed abundances and $\Delta Y$ simultaneously).
Table \ref{tab:data} shows $\Delta Y$, [Fe/H], and the spread in the measurements of [Al/Fe], [N/Fe], and [Na/Fe] between P1 and P2 stars for each of the 26 clusters studied.

The thermohaline mixing that occurs after the RGB bump could potentially  cause problems when analyzing the [C/Fe] and [N/Fe] abundance spreads \citep[e.g.,][]{Gratton2000}.
Since six of the clusters in our sample of 26 only have data for stars above the RGB  bump (i.e., NGC 2808, NGC 5024, NGC 5466, NGC 6388, NGC 6441, and NGC 7089), it is possible that for these clusters the observed abundance dispersion is smaller than that present below the bump\footnote{\href{https://doi.org/10.5281/zenodo.13289055}{https://doi.org/10.5281/zenodo.13289055}.}.

\section{Dilution models}\label{sec:modelo}

This work focuses on testing the basic performance of the enrichment scenarios with AGB, IB, and FRMS stars as polluters. As in the work carried out by \citet{Bastian2015}, the aim is to compare the observed abundances with those predicted by the models, taking into account the maximum difference in He for a given cluster.
Therefore, we made the same assumptions as in that study:
\begin{enumerate}
     \item the P1 (initial) mass fraction of He in all clusters is $Y=0.25$
     \item the polluter yields are diluted in the same way for each cluster.
\end{enumerate}

We start by defining the abundances of the P1 stars for each of the studied elements of a given cluster.
Unlike \citet{Bastian2015}, who assigned these values manually on a case-by-case basis, we analyzed which percentile of the abundance distribution of each element corresponds to the P1 values.
This ensures that the selection criteria for the P1 abundances are the same for each cluster of a given metallicity.
It was decided to use the 5th percentile of the abundance distribution for N, Al, and Na, the 75th for C, the 65th for Mg, and the 80th for O.
To satisfy the second assumption, we take these values as the ``zero-point'' for our models.
We then work out the offsets for the different elements for each  model (polluter yields) and apply them relative to the P1 value (zero point) of a cluster of a given [Fe/H] to model the yields of the polluter at that particular metallicity.

\begin{figure}
	\centering
	\includegraphics[width=\columnwidth]{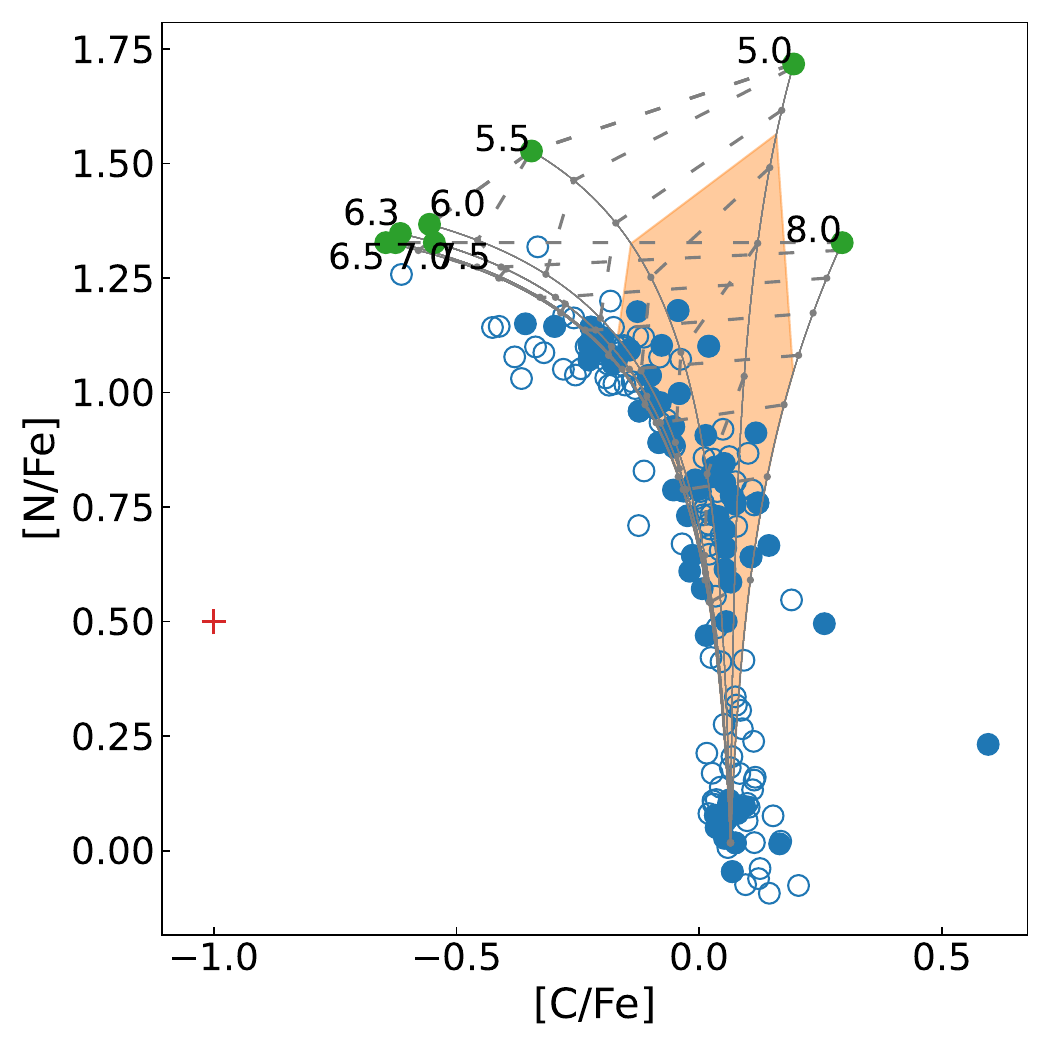}
	\caption{Comparison of the observed [N/Fe] and [C/Fe] abundances in NGC 104 stars with the polluter yields for the AGB scenario. The blue circles correspond to the stars in NGC 104. The filled circles indicate that the star is present in both the APOGEE and the \citet{Carretta2009a,Carretta2009b} datasets, the empty ones indicate that it is present in only one of the datasets. The green circles represent the yields of AGB stars of different masses, labeled with the mass of the parent star in solar masses. Solid gray lines mark the dilution sequence, with the upper and lower ends of $\text{f} = 0$ (undiluted) and $\text{f} = 1$ (fully diluted), respectively. Dashed gray lines connect the points of the dilution curves with the same $Y$, starting at $Y = 0.25$ at the bottom and increasing in nine equally spaced steps toward the top of the plot. The maximum spread of He is by the orange shaded region.
		It represents the range of He values of the cluster's stars; all stars in the cluster should fill this region for the scenario to correctly predict the observations. The red error bars show the mean uncertainty of the measured abundances.}
	\label{fig:parts}
\end{figure}

\begin{figure*}
	\centering
	\begin{subfigure}{\linewidth}
		\centering
		\includegraphics[width=\linewidth]{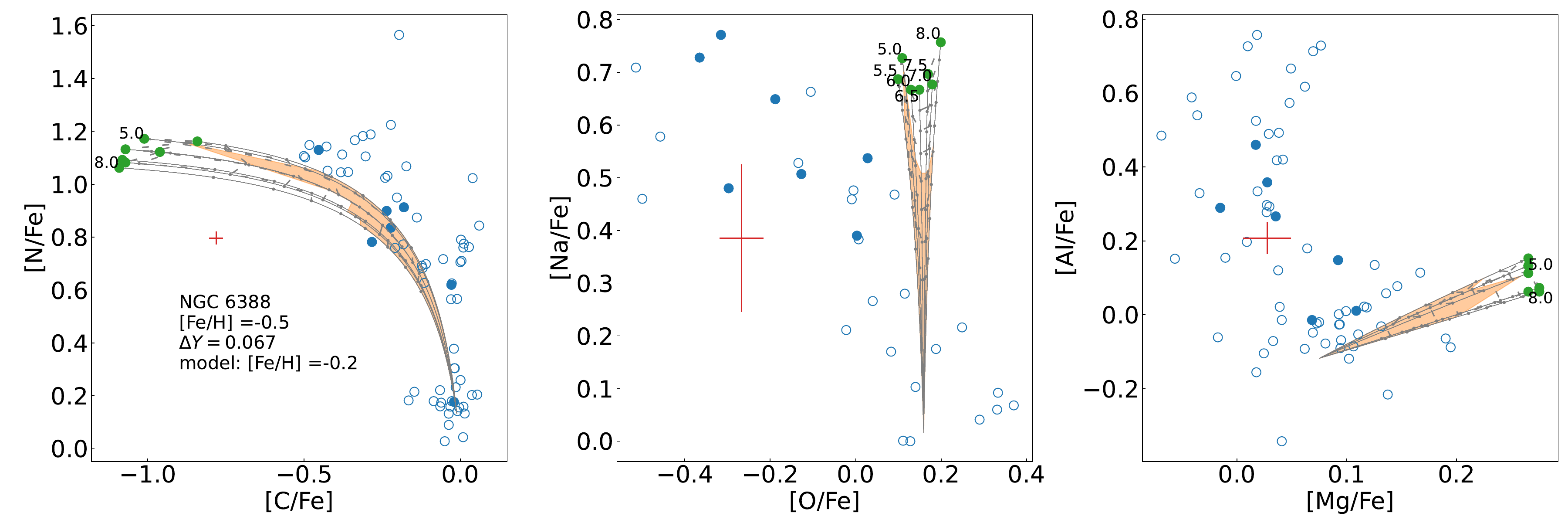}
	\end{subfigure}
	\begin{subfigure}{\linewidth}
		\centering
		\includegraphics[width=\linewidth]{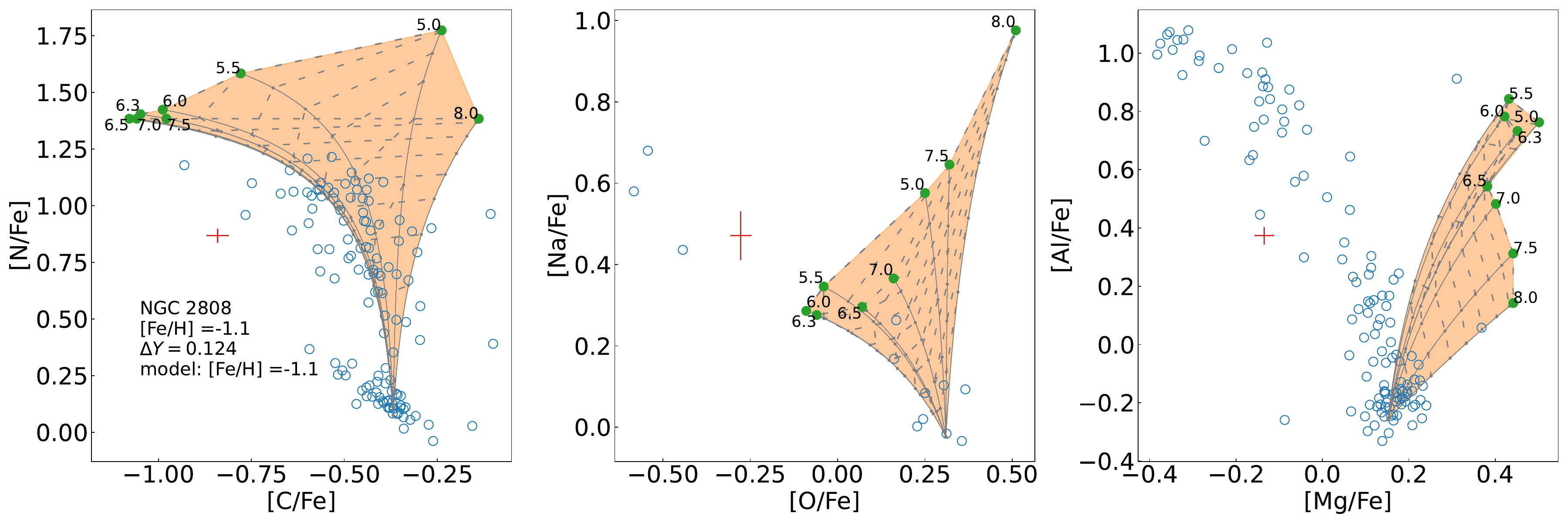}
	\end{subfigure}
	\begin{subfigure}{\linewidth}
		\centering
		\includegraphics[width=\linewidth]{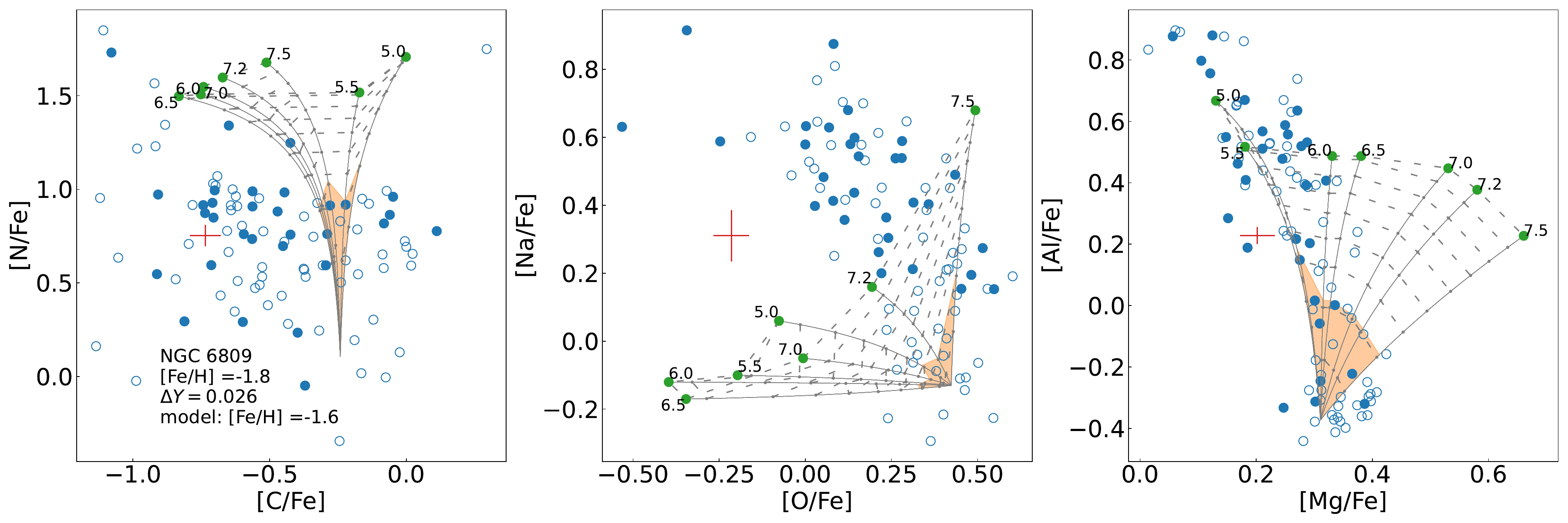}
	\end{subfigure}
	\caption{Results for the AGB scenario for the clusters, from top to bottom: NGC 6388, NGC 2808, and NGC 6809. Similar to Fig. \ref{fig:parts} but for all relevant elements. Each panel shows a different correlation. \textit{Left}: [N/Fe] vs. [C/Fe]. \textit{Middle}: [Na/Fe] vs. [O/Fe]. \textit{Right}: [Al/Fe] vs. [Mg/Fe].}
	\label{fig:AGB}
\end{figure*}

Each scenario presents the yields for its polluters as a single value for each stellar mass.
However, cluster stars show a spread of values for each abundance.
It is then necessary to dilute the polluter yields with the abundance of the first population until we reach the P1 value.
This means that the polluters release their material into the intracluster medium, and as star formation continues, the enriched material is depleted.
As the abundance of He is also effected by this dilution, the He mass fraction must also be taken into account when constructing these models.
To this end, we model the mixing of the pure or undiluted material ejected by the polluters with that of the material with P1 composition (as required by most scenarios), using different mixing fractions, f, via the following formulas for the chemical abundances and the He mass fraction, respectively:
\begin{equation}
    \text{[el/Fe](f)} = \log_{10} \left[(1 -\text{f})\cdot 10^{\text{[el/Fe]}_m} +  \text{f} \cdot 10^{\text{[el/Fe]}} \right],
\end{equation}
\begin{equation}
    Y(\text{f}) = (1 - \mbox{f})\cdot Y_m + \mbox{f}\cdot  Y,
\end{equation}
where [el/Fe]$_m$ and $Y_m$ are the theoretical yields for a given element el and helium, respectively, and [el/Fe] and $Y$ are the P1 values.
Here the mixing fraction represents how diluted the theoretical yields are, 0\% dilution ($\text{f}=0$) represents the maximum values of the theoretical yields and 100\% dilution ($\text{f}=1$) represents the P1 values.
Fig. \ref{fig:parts} shows the comparison of the polluter yields with the observed yields in NGC 104 with the corresponding labels.
The interpretation of this type of plot is presented in the following section.

Finally, we reiterate that this study only tests whether the self-enrichment scenarios can predict the observed abundances in GCs. 
These scenarios require primordial material to dilute the ejecta from the polluter stars.
The origin of this material is still an open debate \citep[e.g.,][]{Conroy2010,DErcole2011,Renzini2015,Bastian2018,Gratton2019} and beyond the scope of this paper.
We choose to assume this material exists, focusing only on a single, yet important, aspect of this complex problem.
It gives these models a ``best case'' scenario which can hopefully be expanded on using the results from this and future investigations.

\begin{figure*}
	\centering
	\includegraphics[width=\linewidth, keepaspectratio]{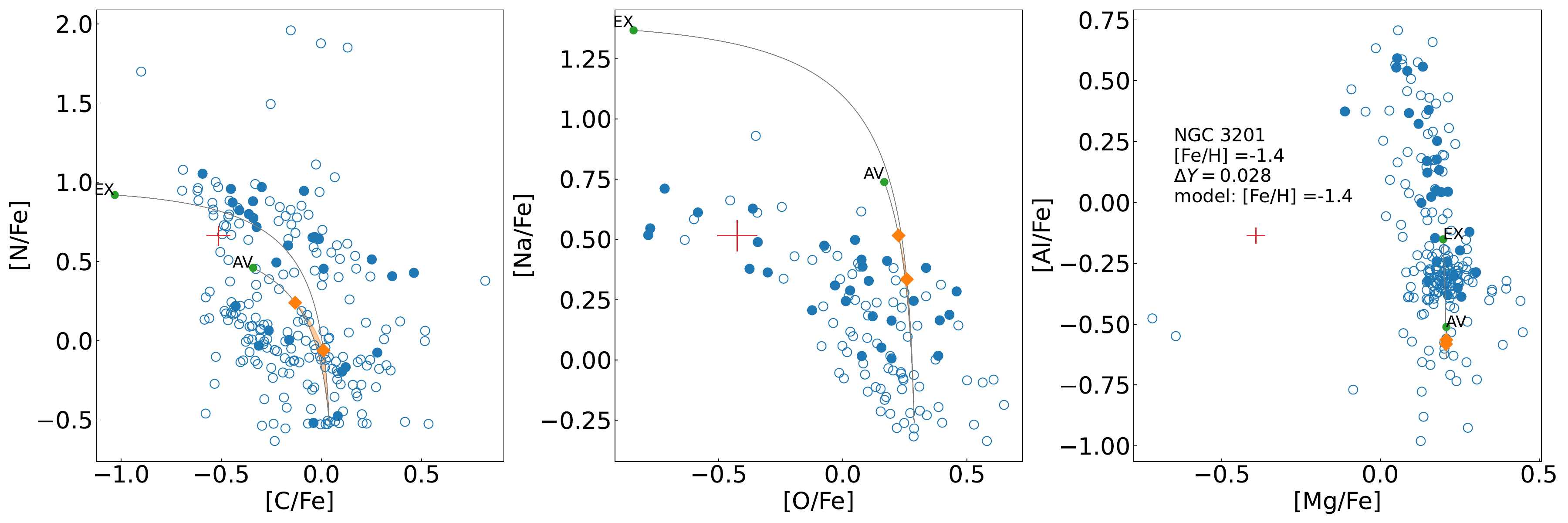}
	\caption{Comparison of the observed abundances in NGC 3201 with the average (AV) and extreme (EX) polluter yields of the IB scenario. \textit{Left}: [N/Fe] vs. [C/Fe]. \textit{Middle}: [Na/Fe] vs. [O/Fe]. \textit{Right}: [Al/Fe] vs. [Mg/Fe].}
	\label{fig:NGC3201IB}
\end{figure*}

\section{Results}

As mentioned above, the aim is to analyze whether the yields of the polluters present in the three self-enrichment scenarios, AGB, IB, and FRMS, proposed to explain the multiple populations present in GCs can explain the observed spread in the abundances of a given pair of elements and how they correlate with each other.
In this analysis, He values are taken as reference and are fixed for the comparisons.
For the abundances of a given polluter to be a good description of the abundances of the cluster stars (blue points in Fig. \ref{fig:parts}), the latter must be distributed within the shaded orange region (i.e., the expected limit for the abundances of the different elements given the observed He spread), as shown in Fig.~\ref{fig:parts}.
A given model underestimates the observed chemical abundances if the blue points in the figure are not constrained within the shaded region and overestimates the observed values if the stars do not reach the limits of the shaded region.
It is important to note that the material ejected by the polluters has the yield indicated by the green dots.
As the stars form, the polluted material is depleted until it reaches the values of P1.

Due to the large number of GCs tested and the similarities between the results, only the most representative cases are shown for each scenario.

\subsection{AGB scenario}

All results for this scenario were obtained using the intermediate mass AGB yields from \citet{Ventura2013}, which primarily produce the correlations seen in GCs.
Since they provided three different metallicities ($Z = 3 \times 10^{-4}, 10^{-3}, 8 \times 10^{-3}$), we compare the different clusters with the yields of the AGB polluters with the nearest metallicity.
Fig. \ref{fig:AGB} shows the results for three clusters, NGC 6388, NGC 2808, and NGC 6809, representing high, middle and low metallicity GCs respectively. 

In general, we can see that this scenario can qualitatively predict the characteristic abundances seen in the GCs. The blue circles in Fig \ref{fig:AGB}, representing the GC stars, follow the gray dilution lines.
However, there are some inconsistencies in the model that were not originally observed by \citet{Bastian2015} due to them only having access to measurements for [Na/Fe] and [O/Fe].
There is a general trend that GC stars with high [Na/Fe] also tend to have high [Al/Fe].
We can see this more effectively in the top panel of Fig. \ref{fig:AGB}, where the solid blue dots in both the middle and right panels have relatively high [Na/Fe] and [Al/Fe].
So, in principle, the polluter that produced the most [Na/Fe] should also have produced the most [Al/Fe].
For the AGB models of [Fe/H] $= -1.1$ and $-1.6$ this is not easy to explain, as at these metallicities, high-mass AGB stars are responsible for the high [Na/Fe].
However, these models predict that the highest [Al/Fe] come from low-mass AGB stars.
Moreover, the models of the most massive AGB stars ($\geq 7.5\text{M}_\sun$), which produce high [Na/Fe] abundances comparable to the observed ones, do not predict a depletion in [O/Fe] and [Mg/Fe], but an enhancement, which is not observed in any case.
Continuing this point, as [Al/Fe] increases we see an enhancement in [Mg/Fe] across all masses and metallicities, save for the lowest masses ($5.0\text{M}_\sun$ and $5.5\text{M}_\sun$) at [Fe/H] $=-1.6$.

N is relatively constant for AGB stars of all masses for a given metallicity, thus both intermediate and high mass AGB stars are consistent with the extreme values of [N/Fe] seen in GCs. However, this is not the case for [C/Fe], as there are some inconsistencies in the mass extremes at intermediate/low metallicities.

More specifically, the AGB yields succeed and fail in different ways depending on the metallicity studied.
At high metallicities, such as in NGC 6388 shown in the top panels of Fig. \ref{fig:AGB}, N is predicted in the proper range, however the value of C is always underestimated.
The model predicts more Mg than observed and little to no change in O or Al.

For the cluster NGC 2808, shown in the middle panels of Fig. \ref{fig:AGB}, we can see that model succeeds in predicting the values of [N/Fe], [C/Fe], and [Na/Fe].
However, it does not satisfactorily predict the abundances of the other elements. The spread for predicted for [O/Fe], [Al/Fe], and [Mg/Fe] are too small for the given He observed.
At this metallicity, we also see that all AGB stars enhance [Al/Fe] and [Mg/Fe] simultaneously, which is not seen in any of the GCs studied.

The comparison for NGC 6809, a low metallicity cluster, shown in the bottom panels of Fig. \ref{fig:AGB}, shows that the dispersion of the measured [C/Fe], [Mg/Fe], and [O/Fe] abundances are larger than the model prediction, since the blue points are not within the shaded orange region, the model seems to underestimate the dispersion of the abundances of these elements.

\begin{figure*}
	\sidecaption
	\includegraphics[width=12cm]{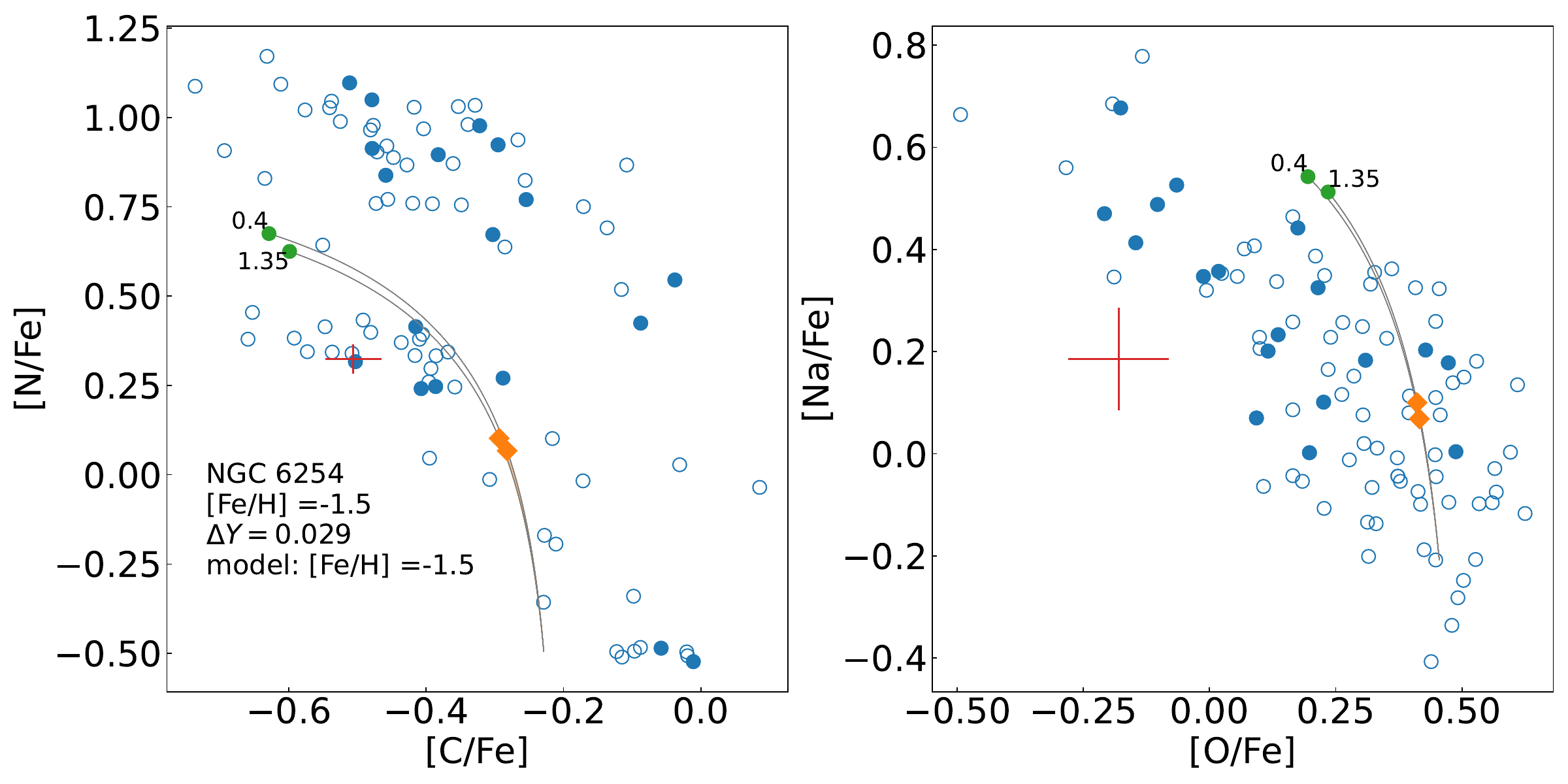}
	\caption{Comparison of the observed abundances in NGC 6254 with the polluter yields of the FRMS scenario. The numbers by the green circles indicate the slope of the IMF of the polluter stars. \textit{Left}: [N/Fe] vs. [C/Fe]. \textit{Right}: [Na/Fe] vs. [O/Fe].}
	\label{fig:NGC6254FRMS}
\end{figure*}

Furthermore, the observed [Na/Fe] and [Al/Fe] values exceed the prediction of the orange region, that is to say, the low-metallicity AGB models underestimates this abundance.
It does, however, correctly predict the spread of [N/Fe], as most of the blue dots are within the P1 value and the maximum value of [N/Fe] permitted by the orange region.
In addition, we observe that only metal-poor ($\mbox{[Fe/H]}\sim-1.6$) low-mass AGB stars ($\sim5-6 M_\sun$) display the observed anti-correlation between [Mg/Fe] and [Al/Fe]. More massive stars, and higher metallicity AGB models predict correlated Al-Mg abundances.

As mentioned before, the Al-Mg behavior of low-mass AGB stars is particularly interesting when comparing the measurements of [Na/Fe] and [O/Fe].
For example, at these metallicities, low-mass AGB stars do not predict a significant [Na/Fe] enhancement with respect to the P1 abundance. Significant [Na/Fe] enhancements are only expected for the material ejected by the most massive AGB stars ($\sim7.5 M_\sun$), but these stars produce an no or only very slight enhancements in [O/Fe], in contrast with observations where the most [Na/Fe] rich stars are the most depleted in [O/Fe].

However, it is not clear how AGBs of this metallicity can do this, as the high mass AGBs produce high [Na/Fe] and the low mass AGBs produce low [O/Fe].
We believe if we were to combine the ejecta of the two we would not reach extreme (high or low) Na-O values.
For this to work, only the Na atoms (but not the O) of the massive stars and the atoms of O (but not the Na) of the low mass star have to come together to produce extreme abundance stars.

The scenario correctly predicts the general trends for these relationships using medium- and high-mass AGB stars. However, when we test these aspects quantitatively, we see that it cannot simultaneously predict the variations in abundances between elements from star to star, since high-mass AGBs do not appear to produce enough Al to account for the observed variations. Furthermore, except in the case of low-mass, low-metallicity AGBs, the yields from \citet{Ventura2013} predict a positive correlation between these two elements instead of the observed negative correlation.

\subsection{IB scenario}

Here we explore the scenario present in \citet{deMink2009}, who propose IBs as the polluters.
Specifically, two stars (15 and 20 $M_\sun$) are used, both with 12 day orbital periods and $Z=5\times 10^{-4}$.
For the discussion in this section we have chosen a cluster of the same metallicity, but the conclusions drawn from the clusters with very different metallicities\footnote{\label{note:all}\href{https://doi.org/10.5281/zenodo.13333930}{https://doi.org/10.5281/zenodo.13333930}} should be taken with a grain of salt.
As with the AGB models, we see that IB model can predict the general trends of the observed data.
Although most stars exceed the abundance represented by the maximum value of He, there is an important subset of stars that follow the gray lines, showing that this scenario can predict the qualitative trends for most clusters, specially in the case of the N-C correlation.

We see in general that the average yields for IBs only show significant changes for [Na/Fe], [N/Fe], and [C/Fe], staying relatively constant for [O/Fe], [Al/Fe], and [Mg/Fe].
The extreme yields, however, fix this problem for [O/Fe].
This is an issue even for [Mg/Fe], whose the relative difference in abundance between P1 and P2 stars is not as great as in other elements.
In most cases the average yields for N, C, and Na, and the extreme yields for O are consistent with the maximum yields present in the cluster.
However, we note that for most GCs, when we take into consideration the measured He value the scenario can only predict the spread in Na, but not any of the other elements studied.
Again, the Al-Mg correlation is worth noting, where we see that [Mg/Fe] remains almost constant for any given value of [Al/Fe] and that even the extreme IB yields cannot account for the spread in Al in most clusters.

Specifically, Fig. \ref{fig:NGC3201IB} shows the comparison for NGC 3201 in this scenario.
As stated, the model underestimates the expected abundances for N, C, Al, Mg, and O for a given value of He, as the observed values are not constrained within the orange region.
On the other hand, the model predicts well the abundance of [Na/Fe], since the observed values reach the orange diamonds.

This scenario manages to predict the qualitative trends for NGC 3201. In other clusters, such as NGC 2808\textsuperscript{\ref{note:all}}, it can correctly predict the N values seen in the cluster.
However, in general the elements N, C, Al, and Mg are underestimated, while the model manages to predict the value of Na and O within the established uncertainties.
It can only achieve this, though, considering the most extreme yields found in the model and even then this does not match the spread in [Al/Fe] and [Mg/Fe] seen in most clusters.

\begin{figure*}
	\centering
	\begin{subfigure}{\linewidth}
		\centering
		\includegraphics[width=\linewidth]{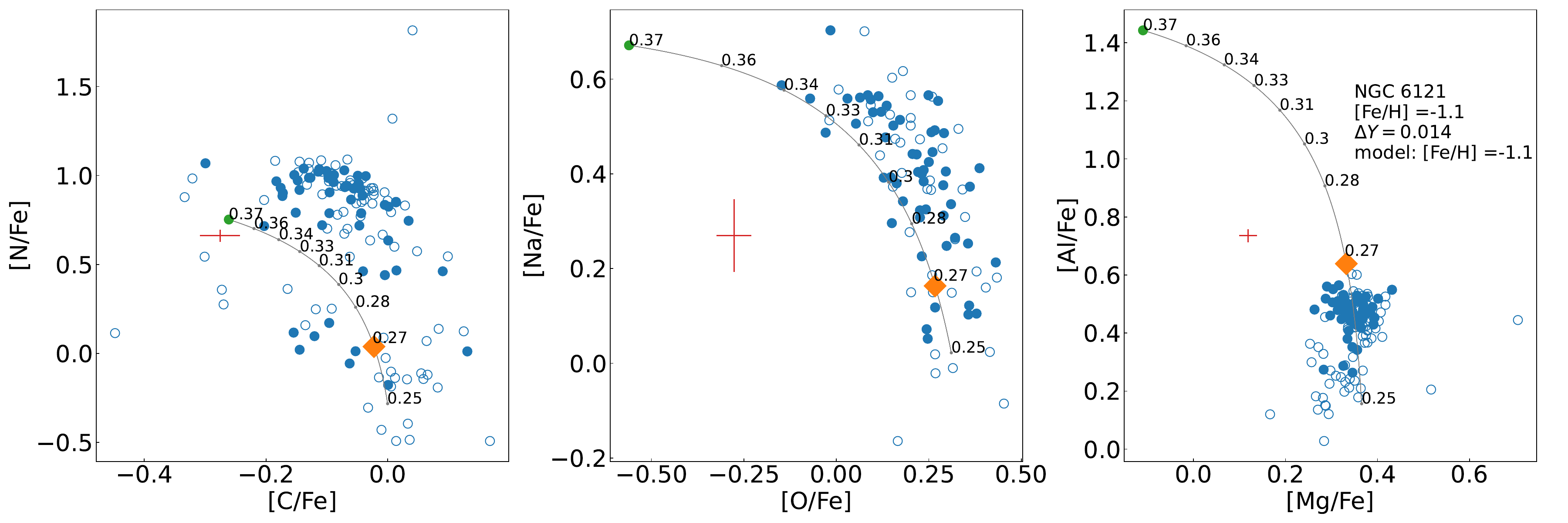}
	\end{subfigure}
	\begin{subfigure}{\linewidth}
		\centering
		\includegraphics[width=\linewidth]{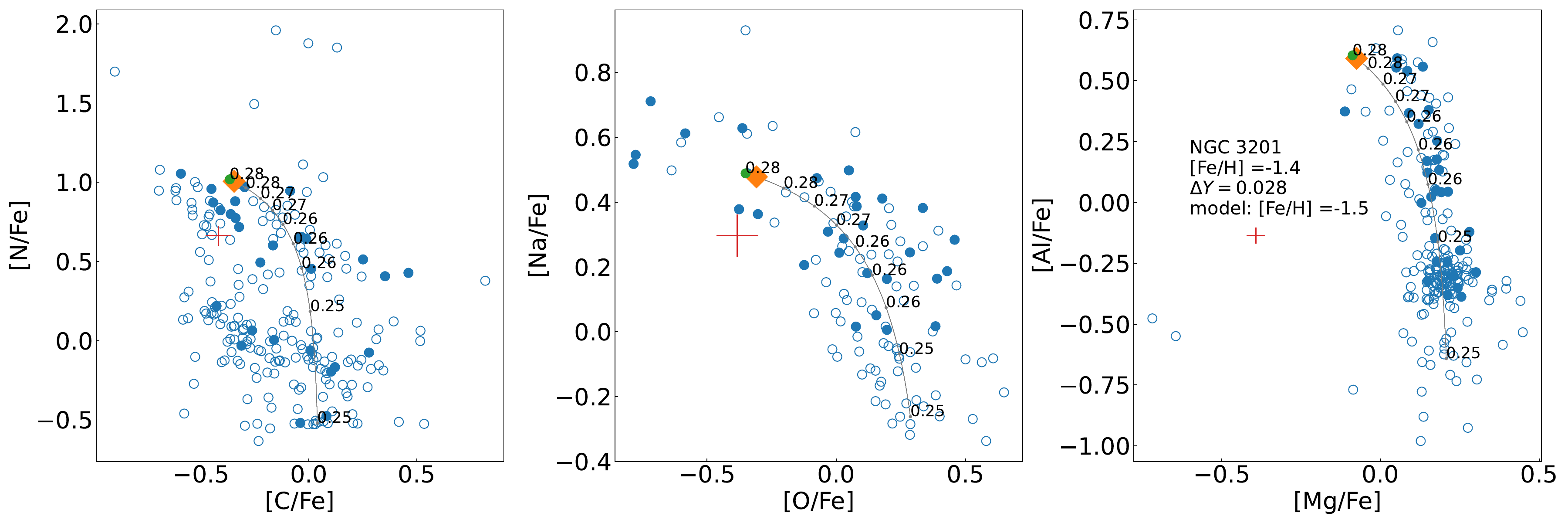}
	\end{subfigure}
	\begin{subfigure}{\linewidth}
		\centering
		\includegraphics[width=\linewidth]{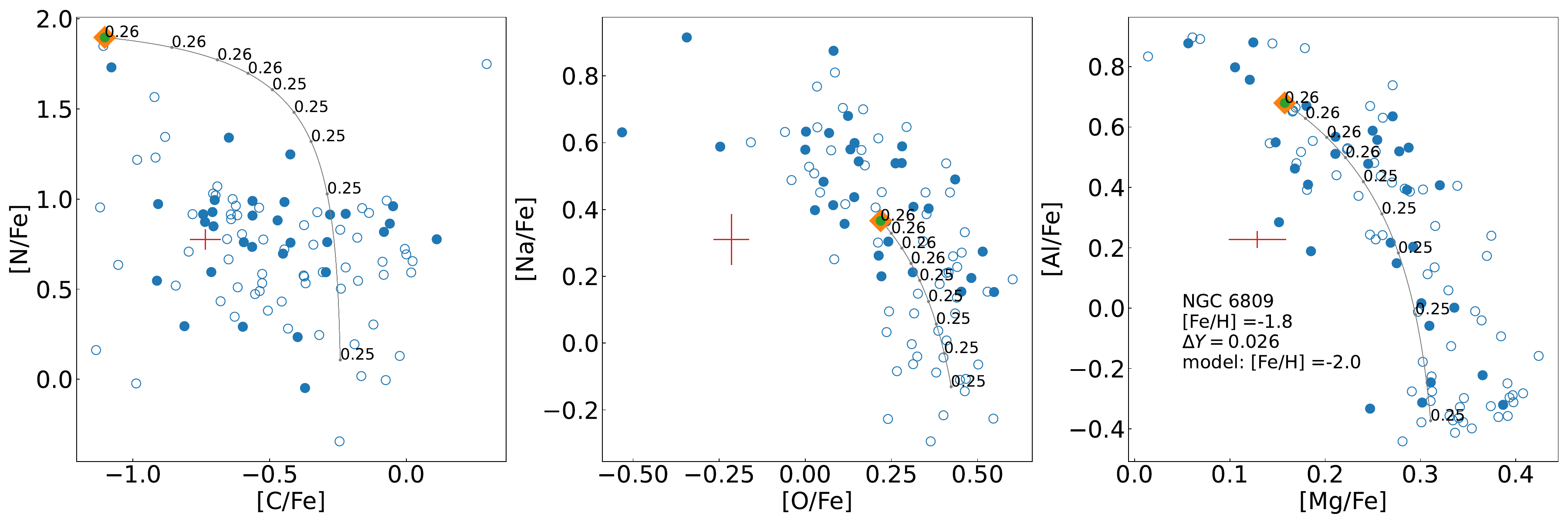}
	\end{subfigure}
	\caption{Results for the empirical model. From top to bottom: NGC 6121, NGC 3201, and NGC 6809. Each panel shows a different correlation, from left to right: [N/Fe] vs. [C/Fe], [Na/Fe] vs. [O/Fe], [Al/Fe] vs. [Mg/Fe].}
	\label{fig:emprical}
\end{figure*}

\subsection{FRMS scenario}

The scenario described by \citet{Decressin2007a,Decressin2007b} uses stars with masses of $20 - 120 M_\sun$, $\frac{\Omega}{\Omega_{crit}}=0.95$
and $Z = 5 \times 10^{-4}$ as polluters.
As in the previous scenario, we have marked the measured maximum $
\Delta Y$ with an orange diamond in Fig. \ref{fig:NGC6254FRMS}.
The green dots that indicate the polluter yields are accompanied by the slope of the initial mass function (IMF) of the stars that produced them.
The data for this model do not include predictions for [Al/Fe], so we cannot analyze the correlation between Al and Mg for this scenario.
As with the IB model, the study only includes yields for stars with one metallicity, so again, the cluster studied in this section corresponds to this metallicity, but results for clusters with very different metallicities should be treated with caution.

For the two correlations studied, [N/Fe] vs. [C/Fe] and [Na/Fe] vs [O/Fe], the general trends of the data agree with the observations.
The stars in the cluster (blue circles) follow the gray dilution lines for the studied elements.
We see that there is not much difference in the yields for clusters with different IMFs (one with a Salpeter-like slope of 1.35 and another with 0.4) and their yields correspond to the extreme values seen in most GCs.
However, the models predict almost no spread in [C/Fe] and [O/Fe] for a given value of He and both [C/Fe] and [Na/Fe] are underestimated. in most GCs.

Fig. \ref{fig:NGC6254FRMS} shows the results for the scenario for NGC 6254.
We see that the spread of the observed measurements is larger than the model predictions for the abundances of [C/Fe] and [O/Fe].
In the case of [N/Fe] and [Na/Fe], the polluter yields match the extreme values seen in the cluster, however as mentioned previously, the observed values exceed the expected value determined by the observed $\Delta Y$.

We also note that other works \citep[e.g.,][]{Tsiatsiou2024,Gormaz-Matamala2024,Nandal2024} analyze the yields for different FRMS in a more general context, using different velocity ratios, different masses, and different metallicities with updated modeling techniques.
We tested the yields in \citet{Nandal2024}, as their results use stars with metallicities relevant to our study, and found that in general the results are in agreement with the ones discussed in this section\footnote{ \href{https://doi.org/10.5281/zenodo.13289206}{https://doi.org/10.5281/zenodo.13289206}.}.

This pattern is repeated for most of the clusters studied.
The predicted values for the four abundances are underestimated, that is to say, the observed values have a larger spread than the spread allowed by the model, although the general trend between the elements is correct.\\

In general all clusters studied follow these trends\textsuperscript{\ref{note:all}}.

\section{Empirical model}

As seen in the last section, none of these types of stars can satisfactorily predict the peculiar abundances observed in GCs.
The theoretical yields are inherently subject to some uncertainty.
To test whether the differences between the observed and theoretical yields are due to the uncertainties in the model, and whether the manifestation of this phenomenon is consistent from cluster to cluster or a fundamental problem in the self-enrichment scenarios, we have adopted an `empirical' model representing a generic source of abundances.
This model takes the `extreme' P2 yields (95th percentile for N, Al, and Na and 5th percentile for C, Mg, and O) of a cluster to be used as the polluter.
Five clusters of different metallicities (NGC 104, NGC 2808, NGC 6254, NGC 6388, and NGC 6397) were used to obtain the corresponding pollutants.
We used these clusters because they have a wide range of abundances at different metallicities.
This allows us to test generic pollutants with a wide range of [Fe/H] ($-0.5. -0.8, -1.1, -1.5, -2.0$).
Again, the clusters were compared to the polluters with the corresponding closest metallicity.
Since this model contains only one dilution line, the maximum value of $\Delta Y$ is represented by an orange diamond.
We show the results for three clusters in Fig. \ref{fig:emprical}.
The selected clusters show the most representative results for the empirical model and have the most data available for them.

In contrast to the theoretical models, we see that the data points always follow the gray lines, as is expected since these models are based on observations.
When taking into account $\Delta Y$, we see that in most cases the empirical model correctly predicts the maximum [Al/Fe].
For the majority of clusters studied, the blue dots do not exceed the orange diamond.
This, however, is not the case for N and Na. In most of the clusters studied, the measured [N/Fe] and [Na/Fe] surpass or fall short of the orange diamond and are underestimated or overestimated, respectively.
It is important to note that the empirical model also shares the same assumptions mentioned in Section \ref{sec:modelo} as the rest of the scenarios presented in this paper.
Taking this into account, these results may tell us that the decoupling of the elements is not necessarily a problem of the theoretical calculation of the yields, as it is also observed in this empirical model.

\begin{figure}
	\centering
	\includegraphics[width=\columnwidth]{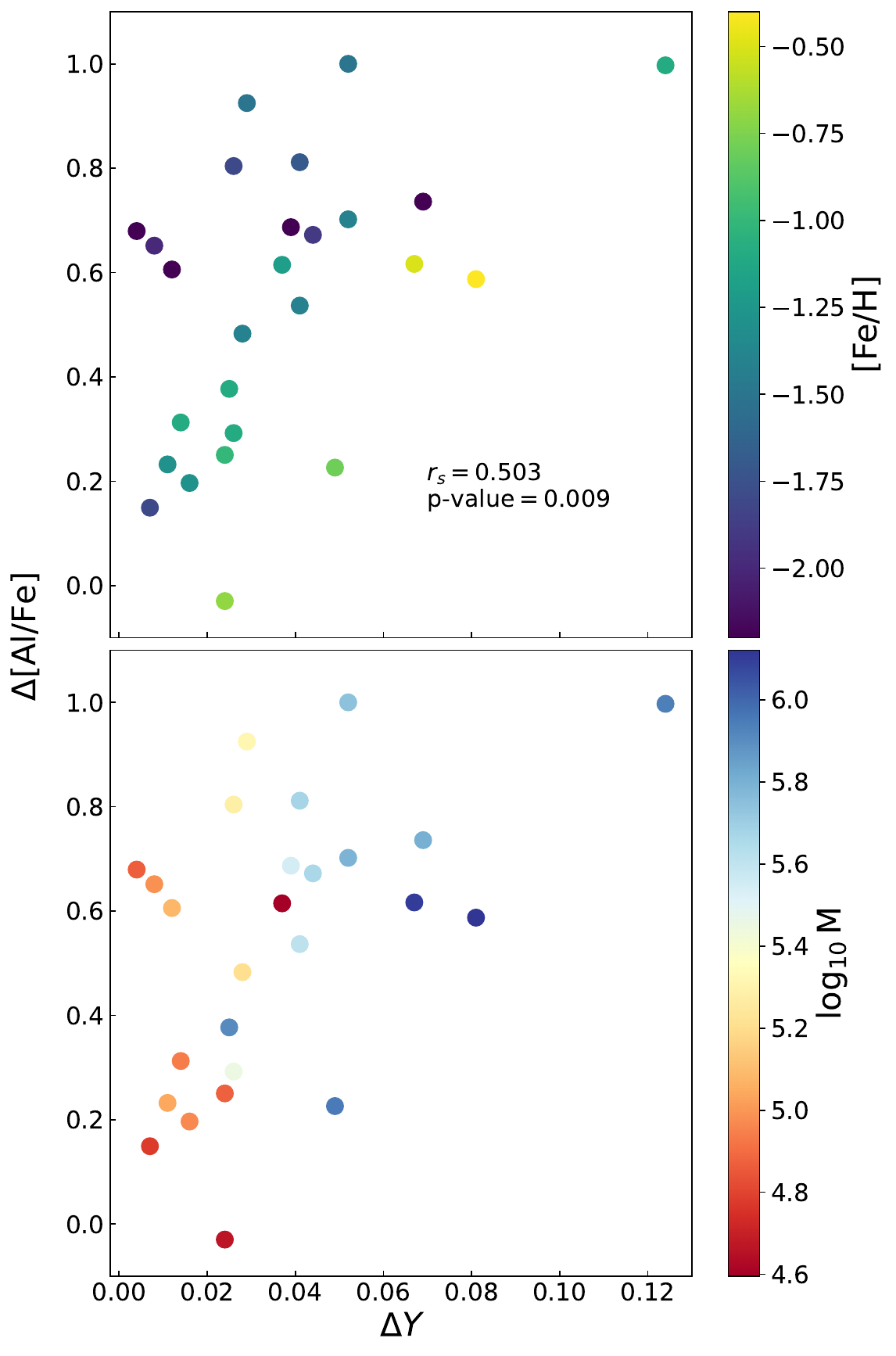}
	\caption{Horizontal axis showing $\Delta Y$ of a given cluster, while the vertical axis shows $\Delta$[Al/Fe] in accordance with equation \eqref{eq:delta}. The values $r_s$ and $p$ are Spearman's correlation coefficient and its p-value respectively. \textit{Top}: the color bar shows the cluster's average metallicity. \textit{Bottom}: the color bar shows the cluster's mass in $M_\sun$.}
	\label{fig:empirico}
\end{figure}

In the top row of Fig. \ref{fig:emprical}, which shows the results for NGC 6121, we see as already mentioned, that the undiluted polluter yields of [N/Fe] and [Na/Fe] match the extreme values of the cluster (green points in Fig. \ref{fig:emprical}), but the cluster stars are not bound to the limit set by the orange diamond (observed $\Delta Y$) and are therefore underestimated.
However, [Al/Fe] is predicted correctly as the stars are bound to limit set by the observed $\Delta Y$.
The middle row of Fig. \ref{fig:emprical} shows NGC 3201. The results for this cluster match the trends mentioned above and correctly predict [N/Fe], [Al/Fe], and [Na/Fe].
Finally, the bottom row of Fig. \ref{fig:emprical} shows cluster NGC 6809. Here, the model correctly predicts both [N/Fe] and [Al/Fe] but underestimates [Na/Fe].

In general, the empirical model cannot predict the distribution of [N/Fe] and [Na/Fe] abundances seen in clusters given the observed $\Delta Y$. However, the model succeeds in predicting the [Al/Fe] distribution of GC stars when $\Delta Y$ is known. These results suggest that the [Al/Fe] abundance variations scale linearly with $\Delta Y$, but not the rest of the CNO elements.

\citet{Carretta2010} found that the interquartile range (IQR) of the ratio between O and Na (O/Na) scales with $M_v$ (mass).
We reproduced the scaling relation present in that study, but do not see a clear correlation between the other elements (i.e., Mg/Al and C/N) and mass. Also, the scaling relation between the O/Na ratio does not translate directly to a scaling relation between just O or just Na with mass.
These comparisons are shown in Appendix \ref{sec:car}.

To verify these results, we compared the $\Delta Y$ with the spread of a given element $\Delta$[el/Fe] of a given cluster, where
\begin{equation}
    \Delta\text{[el/Fe]}=\log\left( 10^{\text{[el/Fe]}_{p95}} - 10^{\text{[el/Fe]}_{p5}} \right),
    \label{eq:delta}
\end{equation}
Here $p95$ and $p5$ represent the 95th and 5th percentile of the cluster's abundance for a given element, respectively.
The results for [N/Fe], [Al/Fe], and [Na/Fe] are listed in Table \ref{tab:data}.

The results for $\Delta Y$ and $\Delta$[Al/Fe] are shown in Fig. \ref{fig:empirico}, where each point in the figure represents a GC. To see if this relationship also correlates with the mass or metallicity of the cluster, we have color-coded the metallicity of the cluster in the top panel and the (log) mass in the bottom panel. We show the Spearman correlation coefficient, $r_s$, between $\Delta$[Al/Fe] and $\Delta Y$ in the top panel of the figure.

From this figure we can see that $\Delta Y$ correlates with $\Delta$[Al/Fe].
This suggests that the enrichment process of GC stars changes the abundances of Al and He in the same way.
That is, if the process increases the amount of He present in the cluster, the amount of Al increases proportionally.
However, this is not the case for other CNO elements such as N and Na (i.e., no significant correlations found between their scatter with $\Delta Y$).
We can also see that there is no trivial correlation between the mass or metallicity in this trend (e.g. at a given $\Delta$[Al/Fe], we find clusters of different masses and [Fe/H]), leading us to believe that the process does not depend strongly on these two factors.

\section{Summary and conclusions}

GCs contain multiple populations of stars with peculiar chemical abundances that cannot be explained by stellar evolution.
To explain this phenomenon, researchers have developed self-enrichment scenarios in which the ejecta of stars belonging to the cluster contaminates the intracluster medium in which the stars form, giving rise to peculiar abundances.
These stars, called polluters, must be able to produce the observed abundance variations and have a short enough lifespan so that the stars in the cluster have negligible age spreads compared to the cluster age (less than a few $\sim10^7$ years).

To find out if the most popular self-enrichment scenarios can reproduce the abundance spread seen in GCs, we tested theoretical predictions for three different types of polluter stars proposed to explain the distinctive chemical patterns found in GCs: AGB, IB, and FRMS stars. We used the latest chemical abundances from the APOGEE survey and precise $\Delta Y$ from HST photometry.
It is important to add here that Na and O were adopted from \citet{Carretta2009a,Carretta2009b}, because Na from APOGEE is not reliable \citep[e.g.,][]{Barbuy2023}.

Our main results are:
\begin{itemize}
    \item The model yields of these stars were able to predict the general trends of the observed abundances for the N-C and Na-O relationships, that is to say, in general the blue dots follow the gray lines predicted by the models in Figs. \ref{fig:AGB} - \ref{fig:NGC6254FRMS} and the in general the clusters tested\textsuperscript{\ref{note:all}}.

    \item The AGB models manage to correctly predict the observed [N/Fe] and [Al/Fe] distributions for 62\% of the clusters, and 67\% of the Na distribution, but only correctly predicted the [C/Fe], [Mg/Fe], and [O/Fe] distributions for 46\%, 26\%, and 27\% of the clusters respectively. 
    
    \item The AGB models also predicts that [Al/Fe] should increase with [Mg/Fe], whereas the observations show the opposite.
    
    \item In general, the models of AGB stars systematically underestimate the maximum of the observed distribution of [Al/Fe] for the high metallicity clusters.
    
    \item The chemical patterns of extreme P2 stars (high in Na and Al, but low in O and Mg) do not follow any of the yields of AGB stars studied in this paper (i.e., either high-mass or low-mass AGB stars). In fact, high-mass AGBs produce high amounts of [Na/Fe] but low amounts of [Al/Fe], while low-mass AGBs produce low [Na/Fe] and high [Al/Fe]. Furthermore, high-mass AGBs predict an enrichment in [O/Fe] and [Mg/Fe], in contrast to the behavior observed in the studied GCs.

    \item Using IBs as the polluters, we found that the observed distribution of [Na/Fe] and [O/Fe] could be reproduced for 67\% of the GCs studied. However, the observed distributions of [N/Fe], [C/Fe], [Al/Fe], and [Mg/Fe] could only describe 35\%, 23\%, 15\%, and 19\% of the clusters respectively.

    \item The FRMS models underestimate the observed distribution of all CNO elements in most GCs.
    The theoretical yields reproduce the observed [N/Fe,] [C/Fe], [Na/Fe], and [O/Fe] distributions of only 35\%, 19\%, 20\%, and 7\%, respectively, of the clusters in our sample.
\end{itemize}

Overall, the theoretical abundances of the AGB, IB, and FRMS polluters do not quantitatively reproduce the observed abundance distributions, although they do match some of the qualitative trends. This is consistent with the results of \citet{Bastian2015}.

We built an empirical model using the abundances of extreme P2 stars as an unspecified source of material polluted with CNO elements. 
The goal is to find if a generic polluter can explain the behavior of the peculiar abundances.
This would show whether this phenomenon is consistent from cluster to cluster and if the inconsistencies between theoretical models and the observations are due to uncertainties in the models or an inherit issue with self enrichment.

The empirical model reproduced well the observed distributions of [Al/Fe] for different clusters, given their $\Delta Y$. However, the empirical model could not reproduce the observed distributions of the other CNO elements. \cite{Bastian2015} reported such decoupling between Na abundance and $\Delta Y$, but no other CNO elements were tested (missing the apparent link between Al and He).

The results from our empirical model suggest that 1) the mechanism responsible for the Al and the He enrichment, increases both in a similar proportion (unlike the other CNO elements), and 2) the decoupling between the predicted Al and He abundances and the rest of the CNO elements is not necessarily a limitation of the theoretical models, as was also found for this "empirical polluter".

Of the various enrichment sources studied here, IBs have considerable potential due to the large parameter space of possible IBs. In this work we have tested the predictions of a specific IB system, so in principle other types of IBs may prove more successful. For example, \citet{Nguyen2024} computed a grid of 204 IB models with primary masses between $10 \text{ and } 40 M_\sun$, periods between 2 and 700 days, and mass ratios between $0.15 \text{ and } 0.9$ at $\text{[Fe/H]}=-1.4$. Unfortunately, exploring such a grid requires a different approach, which is beyond the scope of this paper and will be the focus of a future study. Nevertheless, there are already some positive takeaways from \citet{Nguyen2024}; for example, it has been shown that the mass budget improves by a factor of $\sim 6$ compared to expectations if all stars were single, and the timescale for the return of IB ejecta to the intracluster medium is $\sim 10$ Myr, which is closer to current observational constraints than many other proposed scenarios. However, the models presented in \citet{Nguyen2024} do not overcome one of the main problems found in the \citet{deMink2009} model used in this paper, that is to say, the IBs do not produce the observed high amount of Al enrichment found in many clusters, and can only account for GCs stars with moderate Al enrichment (i.e., a few tenths of dex in [Al/Fe]).

Another type of polluter with interesting potential are very massive stars (VMS, a few $10^2 M_\sun$)\footnote{Not to be confused with super massive stars (SMS) with $10^3 - 10^4 M_\sun$, such as those from \citet{Denissenkov2014} or \citet{Gieles2018}, whose existence remains speculative today.}. \citet{Higgins2023} modeled $50-500 M_\sun$ stars at solar metallicity and discussed their potential to explain the abundance patterns found in GC stars. For example, they find that $>100 M_\sun$ stars are able to enrich their winds with elements processed by hot CNO burning much more than $50 M_\sun$ stars, and the total mass loss is an order of magnitude higher than previously thought. These results go in the right direction to mitigate some of the problems found in the simulations of \citet{Lahen2024}, which used different VMS models \citep[i.e., BoOST,][]{Szecsi2022}.

It is important to emphasize that this work has focused solely on investigating the sources of contamination through the chemical abundances predicted by theoretical (and empirical) models. The assumptions of these models, such as the origin of the primordial material, how the cluster holds onto it, and the resulting age spreads required for the polluters to contaminate the intracluster medium were assumed to be correct in order to determine whether these contaminants could replicate the observed data, thus establishing the most favorable scenario. The results from this paper show that none of the proposed contaminants (even the unspecified empirical source) can accurately replicate the full range of abundances observed in GCs, even under ideal conditions. However, they do correctly represent the general qualitative trends.

These results are expected to open the doors for further studies on the origin of the peculiar abundances in GCs, in particular in models that can incorporate the discovered relationship between He and Al abundances and their decoupling from the abundances of the other CNO elements.

\section*{Acknowledgments}

This study was supported by the Klaus Tschira Foundation.\\

The algorithm was developed in Python 3 \citep{python} using Jupyter Notebooks \citep{Kluyver2016jupyter} and the Astropy \citep{Astropy2013,Astropy2018}, Numpy \citep{harris2020array}, and Matplotlib \citep{Hunter2007} libraries.

\section*{Data Availability}

The observational data used in this investigation is publicly accessible through a third party.
The chemical abundances and metallicities were taken from the Apache Point Observatory Galactic Evolution Experiment (APOGEE) survey \citep{Abdurrouf2022} and \citet{Carretta2009a,Carretta2009b}.
To identify the cluster stars we cross-match the surveys with \citet{Leitinger2023}.
We used the $\Delta Y$ measurements available through \citet{Milone2018}.
The theoretical data for the AGB, IB, and FRMS scenarios where taken from \citet{Ventura2013}, \citet{deMink2009}, and \citet{Decressin2007a,Decressin2007b} respectively.
Finally, we used the MIST Isochrones \citep{dotter_mesa_2016,choi_mesa_2016,paxton_modules_2011,paxton_modules_2013,paxton_modules_2015,paxton_modules_2018} to model the evolutionary tracks of cluster stars.


\bibliographystyle{aa}
\bibliography{referencias}

\begin{thebibliography}{54}
\expandafter\ifx\csname natexlab\endcsname\relax\def\natexlab#1{#1}\fi

\bibitem[{Abdurro’uf {et~al.}(2022)Abdurro’uf, Accetta, Aerts, Aguirre,
  Ahumada, Ajgaonkar, Ak, Alam, Prieto, Almeida, Anders, Anderson, Andrews,
  Anguiano, Aquino-Ortíz, Aragón-Salamanca, Argudo-Fernández, Ata, Aubert,
  Avila-Reese, Badenes, Barbá, Barger, Barrera-Ballesteros, Beaton, Beers,
  Belfiore, Bender, Bernardi, Bershady, Beutler, Bidin, Bird, Bizyaev, Blanc,
  Blanton, Boardman, Bolton, Boquien, Borissova, Bovy, Brandt, Brown,
  Brownstein, Brusa, Buchner, Bundy, Burchett, Bureau, Burgasser, Cabang,
  Campbell, Cappellari, Carlberg, Wanderley, Carrera, Cash, Chen, Chen,
  Cherinka, Chiappini, Choi, Chojnowski, Chung, Clerc, Cohen, Comerford,
  Comparat, da~Costa, Covey, Crane, Cruz-Gonzalez, Culhane, Cunha, Dai, Damke,
  Darling, Jr., Davies, Dawson, Lee, Diamond-Stanic, Cano-Díaz, Sánchez,
  Donor, Duckworth, Dwelly, Eisenstein, Elsworth, Emsellem, Eracleous,
  Escoffier, Fan, Farr, Feng, Fernández-Trincado, Feuillet, Filipp,
  Fillingham, Frinchaboy, Fromenteau, Galbany, García, García-Hernández, Ge,
  Geisler, Gelfand, Géron, Gibson, Goddy, Godoy-Rivera, Grabowski, Green,
  Greener, Grier, Griffith, Guo, Guy, Hadjara, Harding, Hasselquist, Hayes,
  Hearty, Hernández, Hill, Hogg, Holtzman, Horta, Hsieh, Hsu, Hsu, Huber,
  Huertas-Company, Hutchinson, Hwang, Ibarra-Medel, Chitham, Ilha, Imig,
  Jaekle, Jayasinghe, Ji, Johnson, Jones, Jönsson, Katkov, Khalatyan,
  Kinemuchi, Kisku, Knapen, Kneib, Kollmeier, Kong, Kounkel, Kreckel,
  Krishnarao, Lacerna, Lane, Langgin, Lavender, Law, Lazarz, Leung, Leung,
  Lewis, Li, Li, Lian, Liang, Lin, Lin, Lin, Lintott, Long, Longa-Peña,
  López-Cobá, Lu, Lundgren, Luo, Mackereth, de~la Macorra, Mahadevan,
  Majewski, Manchado, Mandeville, Maraston, Margalef-Bentabol, Masseron,
  Masters, Mathur, McDermid, Mckay, Merloni, Merrifield, Meszaros, Miglio,
  Mille, Minniti, Minsley, Monachesi, Moon, Mosser, Mulchaey, Muna, Muñoz,
  Myers, Myers, Nadathur, Nair, Nandra, Neumann, Newman, Nidever, Nikakhtar,
  Nitschelm, O’Connell, Garma-Oehmichen, de~Oliveira, Olney, Oravetz,
  Ortigoza-Urdaneta, Osorio, Otter, Pace, Padilla, Pan, Pan, Parikh, Parker,
  Peirani, Ramírez, Penny, Percival, Perez-Fournon, Pinsonneault, Poidevin,
  Poovelil, Price-Whelan, de~Andrade~Queiroz, Raddick, Ray, Rembold, Riddle,
  Riffel, Riffel, Rix, Robin, Rodríguez-Puebla, Roman-Lopes, Román-Zúñiga,
  Rose, Ross, Rossi, Rubin, Salvato, Sánchez, Sánchez-Gallego, Sanderson,
  Rojas, Sarceno, Sarmiento, Sayres, Sazonova, Schaefer, Schiavon, Schlegel,
  Schneider, Schultheis, Schwope, Serenelli, Serna, Shao, Shapiro, Sharma,
  Shen, Shetrone, Shu, Simon, Skrutskie, Smethurst, Smith, Sobeck, Spoo,
  Sprague, Stark, Stassun, Steinmetz, Stello, Stone-Martinez, Storchi-Bergmann,
  Stringfellow, Stutz, Su, Taghizadeh-Popp, Talbot, Tayar, Telles, Teske,
  Thakar, Theissen, Tkachenko, Thomas, Tojeiro, Toledo, Troup, Trump, Trussler,
  Turner, Tuttle, Unda-Sanzana, Vázquez-Mata, Valentini, Valenzuela,
  Vargas-González, Vargas-Magaña, Alfaro, Villanova, Vincenzo, Wake,
  Warfield, Washington, Weaver, Weijmans, Weinberg, Weiss, Westfall, Wild,
  Wilde, Wilson, Wilson, Wilson, Wolf, Wood-Vasey, Yan, Zamora, Zasowski,
  Zhang, Zhao, Zheng, Zheng, \& Zhu}]{Abdurrouf2022}
Abdurro’uf, Accetta, K., Aerts, C., {et~al.} 2022, \apjs, 259, 35

\bibitem[{{Astropy Collaboration} {et~al.}(2018){Astropy Collaboration},
  Price-Whelan, Sipőcz, Günther, Lim, Crawford, Conseil, Shupe, Craig,
  Dencheva, Ginsburg, VanderPlas, Bradley, Pérez-Suárez, de~Val-Borro,
  Aldcroft, Cruz, Robitaille, Tollerud, Ardelean, Babej, Bach, Bachetti,
  Bakanov, Bamford, Barentsen, Barmby, Baumbach, Berry, Biscani, Boquien,
  Bostroem, Bouma, Brammer, Bray, Breytenbach, Buddelmeijer, Burke, Calderone,
  Rodríguez, Cara, Cardoso, Cheedella, Copin, Corrales, Crichton, D’Avella,
  Deil, Depagne, Dietrich, Donath, Droettboom, Earl, Erben, Fabbro, Ferreira,
  Finethy, Fox, Garrison, Gibbons, Goldstein, Gommers, Greco, Greenfield,
  Groener, Grollier, Hagen, Hirst, Homeier, Horton, Hosseinzadeh, Hu, Hunkeler,
  \v{Z}. Ivezić, Jain, Jenness, Kanarek, Kendrew, Kern, Kerzendorf, Khvalko,
  King, Kirkby, Kulkarni, Kumar, Lee, Lenz, Littlefair, Ma, Macleod,
  Mastropietro, McCully, Montagnac, Morris, Mueller, Mumford, Muna, Murphy,
  Nelson, Nguyen, Ninan, Nöthe, Ogaz, Oh, Parejko, Parley, Pascual, Patil,
  Patil, Plunkett, Prochaska, Rastogi, Janga, Sabater, Sakurikar, Seifert,
  Sherbert, Sherwood-Taylor, Shih, Sick, Silbiger, Singanamalla, Singer,
  Sladen, Sooley, Sornarajah, Streicher, Teuben, Thomas, Tremblay, Turner,
  Terrón, van Kerkwijk, de~la Vega, Watkins, Weaver, Whitmore, Woillez, \&
  Zabalza}]{Astropy2018}
{Astropy Collaboration}, Price-Whelan, A.~M., Sipőcz, B.~M., {et~al.} 2018,
  \aj, 156, 123

\bibitem[{{Astropy Collaboration} {et~al.}(2013){Astropy Collaboration},
  Robitaille, Tollerud, Greenfield, Droettboom, Bray, Aldcroft, Davis,
  Ginsburg, Price-Whelan, Kerzendorf, Conley, Crighton, Barbary, Muna,
  Ferguson, Grollier, Parikh, Nair, Günther, Deil, Woillez, Conseil, Kramer,
  Turner, Singer, Fox, Weaver, Zabalza, Edwards, Bostroem, Burke, Casey,
  Crawford, Dencheva, Ely, Jenness, Labrie, Lim, Pierfederici, Pontzen, Ptak,
  Refsdal, Servillat, \& Streicher}]{Astropy2013}
{Astropy Collaboration}, Robitaille, T.~P., Tollerud, E.~J., {et~al.} 2013,
  \aap, 558, A33

\bibitem[{Barbuy {et~al.}(2023)Barbuy, Friaça, Ernandes, Moura, Masseron,
  Cunha, Smith, Souto, Pérez-Villegas, Souza, Chiappini, Queiroz,
  Fernández-Trincado, da Silva, Santiago, Anders, Schiavon, Valentini,
  Minniti, Geisler, Placco, Zoccali, Schultheis, Nitschelm, Beers, \&
  Razera}]{Barbuy2023}
Barbuy, B., Friaça, A. C.~S., Ernandes, H., {et~al.} 2023, \mnras, 526, 2365

\bibitem[{Bastian {et~al.}(2015)Bastian, Cabrera-Ziri, \&
  Salaris}]{Bastian2015}
Bastian, N., Cabrera-Ziri, I., \& Salaris, M. 2015, \mnras, 449, 3333

\bibitem[{Bastian \& Lardo(2018)}]{Bastian2018}
Bastian, N. \& Lardo, C. 2018, \araa, 56, 83

\bibitem[{Baumgardt {et~al.}(2023)Baumgardt, Hénault-Brunet, Dickson, \&
  Sollima}]{Baumgardt2023}
Baumgardt, H., Hénault-Brunet, V., Dickson, N., \& Sollima, A. 2023, \mnras,
  521, 3991, publisher: OUP ADS Bibcode: 2023MNRAS.521.3991B

\bibitem[{Carretta {et~al.}(2009b)Carretta, Bragaglia, Gratton, \&
  Lucatello}]{Carretta2009b}
Carretta, E., Bragaglia, A., Gratton, R., \& Lucatello, S. 2009b, \aap, 505,
  139

\bibitem[{Carretta {et~al.}(2009a)Carretta, Bragaglia, Gratton, Lucatello,
  Catanzaro, Leone, Bellazzini, Claudi, D’Orazi, Momany, Ortolani, Pancino,
  Piotto, Recio-Blanco, \& Sabbi}]{Carretta2009a}
Carretta, E., Bragaglia, A., Gratton, R., {et~al.} 2009a, \aap, 505, 117

\bibitem[{Carretta {et~al.}(2010)Carretta, Bragaglia, Gratton, Recio-Blanco,
  Lucatello, D'Orazi, \& Cassisi}]{Carretta2010}
Carretta, E., Bragaglia, A., Gratton, R.~G., {et~al.} 2010, \aap, 516, A55

\bibitem[{Cassisi \& Salaris(2020)}]{Cassisi2020}
Cassisi, S. \& Salaris, M. 2020, \aapr, 28, 1

\bibitem[{Charbonnel(2016)}]{Charbonnel2016}
Charbonnel, C. 2016, EAS Publications Series, 80-81, 177

\bibitem[{Choi {et~al.}(2016)Choi, Dotter, Conroy, Cantiello, Paxton, \&
  Johnson}]{choi_mesa_2016}
Choi, J., Dotter, A., Conroy, C., {et~al.} 2016, \apj, 823, 102, publisher:
  {IOP} {ADS} Bibcode: 2016ApJ...823..102C

\bibitem[{Cohen(1978)}]{Cohen1978}
Cohen, J.~G. 1978, \apj, 223, 487

\bibitem[{Conroy \& Spergel(2010)}]{Conroy2010}
Conroy, C. \& Spergel, D.~N. 2010, \apj, 726, 36

\bibitem[{Cottrell \& Costa(1981)}]{Cottrell1981}
Cottrell, P.~L. \& Costa, G. S.~D. 1981, \apj, 245, L79

\bibitem[{de~Mink {et~al.}(2009)de~Mink, Pols, Langer, \& Izzard}]{deMink2009}
de~Mink, S.~E., Pols, O.~R., Langer, N., \& Izzard, R.~G. 2009, \aap, 507, L1

\bibitem[{Decressin {et~al.}(2007b)Decressin, Charbonnel, \&
  Meynet}]{Decressin2007b}
Decressin, T., Charbonnel, C., \& Meynet, G. 2007b, \aap, 475, 859

\bibitem[{Decressin {et~al.}(2007a)Decressin, Meynet, Charbonnel, Prantzos, \&
  Ekström}]{Decressin2007a}
Decressin, T., Meynet, G., Charbonnel, C., Prantzos, N., \& Ekström, S. 2007a,
  \aap, 464, 1029

\bibitem[{Denissenkov \& Hartwick(2014)}]{Denissenkov2014}
Denissenkov, P.~A. \& Hartwick, F. D.~A. 2014, \mnras, 437, L21, publisher: OUP
  ADS Bibcode: 2014MNRAS.437L..21D

\bibitem[{Dotter(2016)}]{dotter_mesa_2016}
Dotter, A. 2016, \apjs, 222, 8, publisher: {IOP} {ADS} Bibcode:
  2016ApJS..222....8D

\bibitem[{D’Antona {et~al.}(2016)D’Antona, Vesperini, D’Ercole, Ventura,
  Milone, Marino, \& Tailo}]{DAntona2016}
D’Antona, F., Vesperini, E., D’Ercole, A., {et~al.} 2016, \mnras, 458, 2122

\bibitem[{D’Ercole {et~al.}(2011)D’Ercole, D’Antona, \&
  Vesperini}]{DErcole2011}
D’Ercole, A., D’Antona, F., \& Vesperini, E. 2011, \mnras, 415, 1304

\bibitem[{Gieles {et~al.}(2018)Gieles, Charbonnel, Krause, Hénault-Brunet,
  Agertz, Lamers, Bastian, Gualandris, Zocchi, \& Petts}]{Gieles2018}
Gieles, M., Charbonnel, C., Krause, M. G.~H., {et~al.} 2018, Monthly Notices of
  the Royal Astronomical Society, 478, 2461, publisher: OUP ADS Bibcode:
  2018MNRAS.478.2461G

\bibitem[{Gormaz-Matamala {et~al.}(2024)Gormaz-Matamala, Cuadra, Ekström,
  Meynet, Curé, \& Belczynski}]{Gormaz-Matamala2024}
Gormaz-Matamala, A.~C., Cuadra, J., Ekström, S., {et~al.} 2024, \aap, 687,
  A290, publisher: EDP Sciences

\bibitem[{Gratton {et~al.}(2019)Gratton, Bragaglia, Carretta, D’Orazi,
  Lucatello, \& Sollima}]{Gratton2019}
Gratton, R., Bragaglia, A., Carretta, E., {et~al.} 2019, \aapr, 27, 136

\bibitem[{Gratton {et~al.}(2012)Gratton, Carretta, \& Bragaglia}]{Gratton2012}
Gratton, R.~G., Carretta, E., \& Bragaglia, A. 2012, \aapr, 20, 50

\bibitem[{Gratton {et~al.}(2000)Gratton, Sneden, Carretta, \&
  Bragaglia}]{Gratton2000}
Gratton, R.~G., Sneden, C., Carretta, E., \& Bragaglia, A. 2000, \aap, 354,
  169, series Title: Astrophysics and Space Science Library

\bibitem[{Grundstrom {et~al.}(2007)Grundstrom, Gies, Hillwig, McSwain, Smith,
  Gehrz, Stahl, \& Kaufer}]{Grundstrom2007}
Grundstrom, E.~D., Gies, D.~R., Hillwig, T.~C., {et~al.} 2007, \apj, 667, 505

\bibitem[{Harris {et~al.}(2020)Harris, Millman, van~der Walt, Gommers,
  Virtanen, Cournapeau, Wieser, Taylor, Berg, Smith, Kern, Picus, Hoyer, van
  Kerkwijk, Brett, Haldane, del R{\'{i}}o, Wiebe, Peterson,
  G{\'{e}}rard-Marchant, Sheppard, Reddy, Weckesser, Abbasi, Gohlke, \&
  Oliphant}]{harris2020array}
Harris, C.~R., Millman, K.~J., van~der Walt, S.~J., {et~al.} 2020, \nat, 585,
  357

\bibitem[{Higgins {et~al.}(2023)Higgins, Vink, Hirschi, Laird, \&
  Sabhahit}]{Higgins2023}
Higgins, E.~R., Vink, J.~S., Hirschi, R., Laird, A.~M., \& Sabhahit, G.~N.
  2023, Monthly Notices of the Royal Astronomical Society, 526, 534, publisher:
  OUP ADS Bibcode: 2023MNRAS.526..534H

\bibitem[{Hunter(2007)}]{Hunter2007}
Hunter, J.~D. 2007, Computing in Science \& Engineering, 9, 90

\bibitem[{Karakas \& Lattanzio(2014)}]{Karakas2014}
Karakas, A.~I. \& Lattanzio, J.~C. 2014, \pasa, 31, e030

\bibitem[{Kluyver {et~al.}(2016)Kluyver, Ragan-Kelley, P{\'e}rez, Granger,
  Bussonnier, Frederic, Kelley, Hamrick, Grout, Corlay, Ivanov, Avila, Abdalla,
  \& Willing}]{Kluyver2016jupyter}
Kluyver, T., Ragan-Kelley, B., P{\'e}rez, F., {et~al.} 2016, in Positioning and
  Power in Academic Publishing: Players, Agents and Agendas, ed. F.~Loizides \&
  B.~Schmidt, IOS Press, 87 -- 90

\bibitem[{Kudryashov \& Tutukov(1988)}]{Kudryashov1988}
Kudryashov, A.~D. \& Tutukov, A.~V. 1988, Astronomicheskii Tsirkulyar, 11

\bibitem[{Lahén {et~al.}(2024)Lahén, Naab, \& Szécsi}]{Lahen2024}
Lahén, N., Naab, T., \& Szécsi, D. 2024, \mnras, 530, 645

\bibitem[{Leitinger {et~al.}(2023)Leitinger, Baumgardt, Cabrera-Ziri, Hilker,
  \& Pancino}]{Leitinger2023}
Leitinger, E., Baumgardt, H., Cabrera-Ziri, I., Hilker, M., \& Pancino, E.
  2023, \mnras, 520, 1456

\bibitem[{Maeder \& Meynet(2006)}]{Maeder2006}
Maeder, A. \& Meynet, G. 2006, \aap, 448, L37

\bibitem[{Martocchia {et~al.}(2018)Martocchia, Niederhofer, Dalessandro,
  Bastian, Kacharov, Usher, Cabrera-Ziri, Lardo, Cassisi, Geisler, Hilker,
  Hollyhead, Kozhurina-Platais, Larsen, Mackey, Mucciarelli, Platais, \&
  Salaris}]{Martocchia2018}
Martocchia, S., Niederhofer, F., Dalessandro, E., {et~al.} 2018, \mnras, 477,
  4696, aDS Bibcode: 2018MNRAS.477.4696M

\bibitem[{Milone {et~al.}(2018)Milone, Marino, Renzini, D’Antona, Anderson,
  Barbuy, Bedin, Bellini, Brown, Cassisi, Cordoni, Lagioia, Nardiello,
  Ortolani, Piotto, Sarajedini, Tailo, van~der Marel, \&
  Vesperini}]{Milone2018}
Milone, A.~P., Marino, A.~F., Renzini, A., {et~al.} 2018, \mnras, 481, 5098

\bibitem[{Nandal {et~al.}(2024)Nandal, Meynet, Ekström, Moyano, Eggenberger,
  Choplin, Georgy, Farrell, \& Maeder}]{Nandal2024}
Nandal, D., Meynet, G., Ekström, S., {et~al.} 2024, \aap, 684, A169,
  publisher: EDP Sciences

\bibitem[{Nguyen \& Sills(2024)}]{Nguyen2024}
Nguyen, M. \& Sills, A. 2024, \apj, 969, 18, publisher: IOP Publishing

\bibitem[{Paxton {et~al.}(2011)Paxton, Bildsten, Dotter, Herwig, Lesaffre, \&
  Timmes}]{paxton_modules_2011}
Paxton, B., Bildsten, L., Dotter, A., {et~al.} 2011, \apjs, 192, 3, publisher:
  {IOP} {ADS} Bibcode: 2011ApJS..192....3P

\bibitem[{Paxton {et~al.}(2013)Paxton, Cantiello, Arras, Bildsten, Brown,
  Dotter, Mankovich, Montgomery, Stello, Timmes, \&
  Townsend}]{paxton_modules_2013}
Paxton, B., Cantiello, M., Arras, P., {et~al.} 2013, \apjs, 208, 4, publisher:
  {IOP} {ADS} Bibcode: 2013ApJS..208....4P

\bibitem[{Paxton {et~al.}(2015)Paxton, Marchant, Schwab, Bauer, Bildsten,
  Cantiello, Dessart, Farmer, Hu, Langer, Townsend, Townsley, \&
  Timmes}]{paxton_modules_2015}
Paxton, B., Marchant, P., Schwab, J., {et~al.} 2015, \apjs, 220, 15, publisher:
  {IOP} {ADS} Bibcode: 2015ApJS..220...15P

\bibitem[{Paxton {et~al.}(2018)Paxton, Schwab, Bauer, Bildsten, Blinnikov,
  Duffell, Farmer, Goldberg, Marchant, Sorokina, Thoul, Townsend, \&
  Timmes}]{paxton_modules_2018}
Paxton, B., Schwab, J., Bauer, E.~B., {et~al.} 2018, \apjs, 234, 34, publisher:
  {IOP} {ADS} Bibcode: 2018ApJS..234...34P

\bibitem[{Prantzos \& Charbonnel(2006)}]{Prantzos2006}
Prantzos, N. \& Charbonnel, C. 2006, \aap, 458, 135

\bibitem[{Prantzos {et~al.}(2017)Prantzos, Charbonnel, \&
  Iliadis}]{Prantzos2017}
Prantzos, N., Charbonnel, C., \& Iliadis, C. 2017, \aap, 608, A28

\bibitem[{Renzini {et~al.}(2015)Renzini, D’Antona, Cassisi, King, Milone,
  Ventura, Anderson, Bedin, Bellini, Brown, Piotto, der Marel, Barbuy,
  Dalessandro, Hidalgo, Marino, Ortolani, Salaris, \& Sarajedini}]{Renzini2015}
Renzini, A., D’Antona, F., Cassisi, S., {et~al.} 2015, \mnras, 454, 4197

\bibitem[{Szécsi {et~al.}(2022)Szécsi, Agrawal, Wünsch, \&
  Langer}]{Szecsi2022}
Szécsi, D., Agrawal, P., Wünsch, R., \& Langer, N. 2022, Astronomy \&
  Astrophysics, 658, A125, publisher: EDP Sciences

\bibitem[{Tsiatsiou {et~al.}(2024)Tsiatsiou, Sibony, Nandal, Sciarini, Hirai,
  Ekström, Farrell, Murphy, Choplin, Hirschi, Chiappini, Liu, Bromm, Groh, \&
  Meynet}]{Tsiatsiou2024}
Tsiatsiou, S., Sibony, Y., Nandal, D., {et~al.} 2024, \aap, 687, A307,
  publisher: EDP Sciences

\bibitem[{Van~Rossum \& Drake(2009)}]{python}
Van~Rossum, G. \& Drake, F.~L. 2009, Python 3 Reference Manual (Scotts Valley,
  CA: CreateSpace)

\bibitem[{Vandenberg {et~al.}(2013)Vandenberg, Brogaard, Leaman, \&
  Casagrande}]{Vandenberg2013}
Vandenberg, D.~A., Brogaard, K., Leaman, R., \& Casagrande, L. 2013, \apj, 775,
  134

\bibitem[{Ventura {et~al.}(2013)Ventura, Di~Criscienzo, Carini, \&
  D’Antona}]{Ventura2013}
Ventura, P., Di~Criscienzo, M., Carini, R., \& D’Antona, F. 2013, \mnras,
  431, 3642

\end{thebibliography}



\appendix

\section{Scaling relation between element ratios and mass}\label{sec:car}

\begin{figure}
    \centering
    \includegraphics[width=0.9\columnwidth]{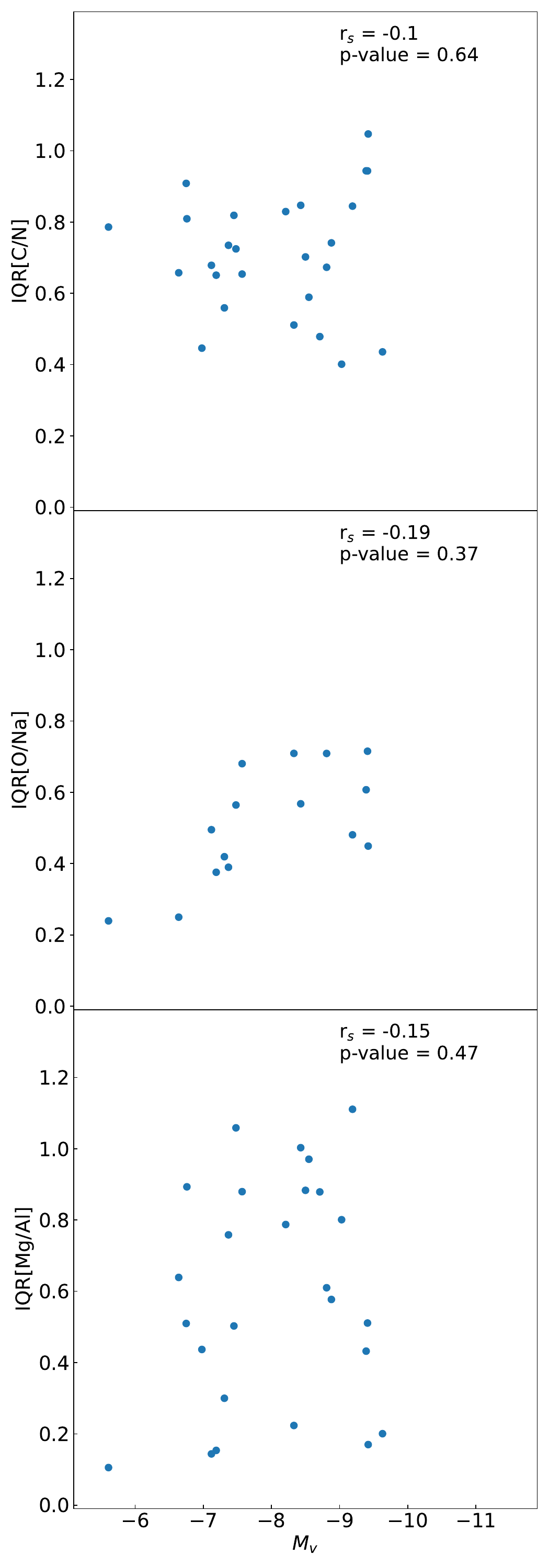}
    \caption{Comparison of the IQR of different ratios between elements versus the cluster mass. From top to bottom, we provide [C/N], [O/Na], and [Mg/Al].}
    \label{fig:check}
\end{figure}

Fig. \ref{fig:check} shows the results of comparing the IRQ of different element ratios to cluster mass, similar to \citet{Carretta2010}.

\end{document}